\documentclass[11pt]{article} \usepackage{mystyle-new}
\usepackage{epsfig,amsmath} \usepackage{graphics}
\usepackage{cite,./mcite}

  \newenvironment{defl}[1]%
  {\begin{list}{}{\settowidth{\labelwidth}{#1}%
  \setlength{\leftmargin}{\labelwidth}%
  \addtolength{\leftmargin}{\labelsep}%
  \setlength{\itemsep}{0pt plus 1pt}
  \setlength{\parsep}{0pt plus 1pt}
  \setlength{\topsep}{0pt plus 1pt}
  \setlength{\partopsep}{0pt plus 1pt}
  \setlength{\parskip}{2mm plus 1mm minus 1mm}
  }}%
  {\end{list}}
\def\lsim{\mathrel{\rlap{\lower4pt\hbox{\hskip1pt$\sim$}}
    \raise1pt\hbox{$<$}}}                
\def\gsim{\mathrel{\rlap{\lower4pt\hbox{\hskip1pt$\sim$}}
    \raise1pt\hbox{$>$}}}                

\newcommand{\cA}{{\cal A}}
\newcommand{\as}{\alpha_\mathrm{s}}

\newcommand{\Pmax}{p}

\def\kt{\ensuremath{k_t}}

\newcommand{\SMALLXC}{SMALLXa,SMALLXb}

\newcommand{\alphasb}{\bar{\alpha}_s}

\def\CASCADE{{\sc Cascade}}

\def\ccfmupdf{{CCFMuPDF}} 
\def\ccfmupdf{{uPDFevolv}} 
\def\TMDplotter{{\sc TMDplotter}}

\makeindex
\begin{document}
\begin{flushright}
DESY 14-060 \\
RAL-P-2014-010 \\
July 2014
\end{flushright}
\begin{center} {\sffamily\Large\bfseries
The CCFM uPDF evolution \\ \vspace*{0.15cm}
{uPDFevolv}  \\ \vspace*{0.15cm}
Version ~1.0.00
\ }
 \\ \vspace{0.5cm}
{ \Large F.~Hautmann$^{1,2,3}$,
H.~Jung$^{4,5}$,
S.~Taheri~Monfared$^{6}$
}\\ \vspace*{0.15cm}
{\large $^1$Dept.\  of Physics and Astronomy, University of   Sussex,   Brighton   BN1  9QH} \\
{\large $^2$ Rutherford Appleton Laboratory,  Chilton  OX11  0QX} \\
{\large $^3$Dept.\  of  Theoretical Physics, University of Oxford,    Oxford OX1 3NP}  \\
{\large $^4$DESY, Hamburg, FRG}\\
{\large $^5$University of Antwerp, Antwerp, Belgium}\\
{\large $^6$School of Particles and Accelerators, Institute for Research in Fundamental Sciences (IPM), P.O.Box 19395-5531, Tehran, Iran}

\end{center}
\begin{abstract}
\ccfmupdf\ is an evolution code for TMD parton densities using the CCFM evolution equation.
A description of the underlying theoretical model and technical realisation is given together with
a detailed program description, with emphasis on parameters
the user  may  want to change.
\end{abstract}
{\sffamily\large\bfseries PROGRAM SUMMARY} \\ \\
{\em Title of Program:} \ccfmupdf\ ~1.0.00
\ \\
{\em Computer for which the program is designed and others on which it is
operable:}   any with standard Fortran 77 (gfortran) and C++, tested on
                 Linux, MAC\\
{\em Programming Language used:}  FORTRAN 77, C++ \\
{\em High-speed storage required:}  No \\
{\em Separate documentation available: } No \\
{\em Keywords: } QCD, small $x$, high-energy factorization, $k_t$-factorization, CCFM, unintegrated PDF (uPDF), transverse momentum dependent PDF (TMD) \\
{\em Nature of physical problem:}
At high energies collisions of hadrons are 
described by parton densities dependent on the longitudinal momentum fraction $x$, the transverse momentum $\kt$ and the evolution scale $\Pmax$  (transverse momentum dependent (TMD) or unintegrated parton density functions (uPDF)). The evolution of the parton density with the scale $\Pmax$ valid at
both small and moderate $x$ is given by the CCFM evolution equation
\\
{\em Method of solution:} 
Since the CCFM evolution equation cannot be solved analytically, a Monte Carlo approach is applied, simulating at each step of the evolution the full four-momenta of the initial state partonic cascade.
\\ 
{\em Restrictions on the complexity of the problem:}   None
\\
{\em Other Program used:}  {\sc Root} for plotting the result. \\
{\em Download of the program:} \verb+https://updfevolv.hepforge.org+\\
{\em Unusual features of the program:}   None \\
\newpage

\section{Theoretical Input}
\subsection{CCFM evolution equation and Transverse Momentum Dependent PDFs}
\label{sec:CCFMEquation}

QCD  calculations of multiple-scale processes  and complex final-states
require  in general   transverse-momentum dependent (TMD),  or
unintegrated,     parton  density and parton decay
 functions~\cite{Collins:2011zzd,Aybat:2011zv,Buffing:2013eka,Buffing:2013kca,Buffing:2012sz,Mulders:2008tf,Jadach:2009gm,Hautmann:2009zzb,Hautmann:2012pf,Hautmann:2007gw}.  
TMD factorization has been proven recently~\cite{Collins:2011zzd} for inclusive and semi-inclusive deep-inelastic scattering (DIS). For special
processes in hadron-hadron scattering, like heavy flavor or heavy  boson (including Higgs) production,
TMD factorization holds in the high-energy limit (small $x$) \cite{Catani:1990eg,Catani:1993ww,Hautmann:2002tu}.
 
In the  framework of high-energy factorization~\cite{Catani:1990xk,Catani:1990eg}
   the deep-inelastic   scattering
cross section can be written as a  convolution in
both longitudinal and transverse momenta  of
the  TMD  parton density function
${\cal  A}\left(x,\kt,\mu\right)$   
 with   off-shell partonic
matrix elements, as follows
\begin{equation}
 \sigma_j  ( x , Q^2 )  = \int_x^1 
d z  \int d^2k_t \
\hat{\sigma}_j( x   ,  Q^2 ,  {    z}   ,  k_t ) \
 {\cal  A}\left( { z} ,\kt,  \Pmax \right)  , 
\label{kt-factorisation}
\end{equation}
with the DIS cross sections
$\sigma_j$ ($j= 2 , L$) related to the  structure functions $F_2$ and $F_L$  by
$\sigma_j = 4 \pi^2 F_j / Q^2$.
The  hard-scattering kernels ${\hat \sigma}_j$ of Eq.~(\ref{kt-factorisation})    
are $k_t$-dependent and the evolution  of the
transverse momentum dependent gluon density
${\cal  A} $ is obtained by combining    the resummation of  small-$x$ logarithmic
contributions~\cite{Lipatov:1996ts,Fadin:1975cb,Balitsky:1978ic}   with   medium-$x$  and large-$x$
 contributions to parton  splitting~\cite{Gribov:1972ri,Altarelli:1977zs,Dokshitzer:1977sg} according to the
 CCFM  evolution equation~\cite{Ciafaloni:1987ur,Catani:1989sg,Marchesini:1994wr}.
 
The factorization formula (\ref{kt-factorisation}) 
allows one to  resum 
logarithmically enhanced $ x \to 0 $ contributions 
  to all orders in perturbation theory, 
both in the  hard
scattering coefficients and
in  the parton evolution,  taking fully into account the
dependence on the factorization scale $\Pmax$ and on the
factorization scheme~\cite{Catani:1994sq,Catani:1993rn}.  

 The  
CCFM evolution
equation~\cite{Ciafaloni:1987ur,Catani:1989sg,Marchesini:1994wr}
 is an exclusive  equation for final state partons
and  includes  finite-$x $
contributions to parton splitting.  It  incorporates
soft gluon coherence for any value of $x$.

\subsubsection{Gluon distribution}
  The evolution equation for the TMD gluon density $\cA(x,\kt,\Pmax)$, depending on
$x$, $\kt$ and the evolution variable $\Pmax$, is 
\begin{eqnarray}
\label{uglurepr1}
  {\cal A} ( x , \kt , p  ) & = &
  {\cal A}_0 ( x , \kt , p  ) +
\int { \frac{dz}{z}} \int { \frac{ d q^2}{q^2}} \
\Theta   (p - z  q) 
\nonumber\\
& \times &
 \Delta_s    (p , z  q)
P ( z, q, \kt)  
\   {\cal A}
 \left( \frac{x}{z} , \kt  + (1-z) q, q \right)
 \hspace*{0.3 cm} ,        
\end{eqnarray}
where $z$ is the longitudinal momentum fraction, $q$ is the angular variable and
the $\Theta$ function specifies the ordering condition of the evolution~\cite{Hautmann:2008vd}.

The first term in the right hand side of Eq.~(\ref{uglurepr1})
is the contribution of the
non-resolvable branchings between  the starting scale
$q_0$ and  the evolution scale $p$,
 and is given by
\begin{equation}
\label{uglurepr2}
  {\cal A}_0 ( x , \kt , p  ) =  {\cal A}_0 ( x , \kt , q_0 )
 \ \Delta_s (p , q_0)
   ,  
\end{equation}
where $\Delta_s$ is the Sudakov form factor, and
 ${\cal A}_0 ( x , \kt , q_0 )$     is the
 starting  distribution  
at scale $q_0$.
The integral term in the right hand side of Eq.~(\ref{uglurepr1})
gives the \kt-dependent branchings in terms of the
 Sudakov form factor $\Delta_s$ and unintegrated
  splitting function $P$.
The  Sudakov form factor $\Delta_s$ is given by 
\begin{equation}
\Delta_s(\Pmax,q_0) =\exp{\left(
 - \int_{q_0^2} ^{\Pmax^2}
 \frac{d q^{2}}{q^{2}}
 \int_0^{1-q_0/q} dz \ \frac{\alphasb(q^2(1-z)^2)}{1-z}
  \right)} ,
  \label{Sudakov-delta}
\end{equation}
with ${\overline \alpha}_s=  C_A \alpha_s / \pi =3 \alpha_s  / \pi$.

For application in Monte Carlo event generators, like \CASCADE\ ~\cite{Jung:2000hk,Jung:2010si}, it is of advantage to write the CCFM evolution equation in differential form:

\begin{equation}
\Pmax^2\frac{d\; }{d \Pmax^2}
   \frac{x \cA(x,\kt,\Pmax)}{\Delta_s(\Pmax,q_0)}=
   \int dz  \   \frac{d\phi}{2\pi}\,
   \frac{{P} (z,\Pmax/z,\kt)}{\Delta_s(\Pmax,q_0)}\,
 x'\cA(x',\kt',\Pmax/z) ,
\label{CCFM_differential}
\end{equation}
where the splitting variable  $x'$  is given by
$x'=x/z$,  ${\kt}' = q_t (1-z)/z + {\kt}$,  and $\phi$ is the
 azimuthal angle  of $q_t$.

For the evolution of the parton densities, however, a forward evolution approach, starting from the low scale $q_0$ towards the hard scale $p$, is used.

The  splitting function $P_{gg}(z_i,q_i,k_{ti})$ for branching $i$
is given by~\cite{Hansson:2003xz}  (set by  \verb+Ipgg=1+, \verb+ns=1+ in \ccfmupdf )
\begin{eqnarray}
P_{gg} (z_i,q_i,k_{ti})
& = & \alphasb(q^2_{i}(1-z_i)^2)  \  \left(
 \frac{ 1 } {1-z_i}  - 1   + \frac{ z_i (1 - z_i)  } {2 }   \right) 
\nonumber\\
&  +  &
\alphasb(k^2_{ti})  
 \ \left(   \frac{1}{ z_i}   - 1   + \frac{ z_i (1 - z_i)  }{  2 }
\right)   \   \Delta_{ns}(z_i,q^2_{i},k^2_{ti}) 
\label{Pgg1}
\end{eqnarray}
where  $\Delta_{ns}$ is  
the non-Sudakov form factor   
defined by
\begin{equation}
\log\Delta_{ns} =  -\alphasb(k^2_{ti})
                  \int_0^1 dz' \left(   \frac{1}{ z'}   - 1   + \frac{ z'(1 - z')  }{  2 } \right)
                        \int \frac{d q^2}{q^2}
              \Theta(k_{ti}-q)\Theta(q-z'q_{ti}).
                  \label{non_sudakov}                  
\end{equation}

In addition to the full splitting function, simplified versions are
useful in applications and are made available.  One  
uses only the singular parts of the splitting function (set
by  \verb+Ipgg=0+, \verb+ns=0+ in \ccfmupdf ): 
\begin{eqnarray}
P_{gg}(z,q,\kt ) & = & \frac{\alphasb(q^2)}{1-z} +
\frac{\alphasb(\kt^2)}{z} \Delta_{ns}(z,q^2,\kt)
\label{Pgg0}
\end{eqnarray}
with
\begin{equation}
\log\Delta_{ns} =  -\alphasb(k^2_{ti})
                  \int_0^1 \frac{dz'}{ z'}                          \int \frac{d q^2}{q^2}
              \Theta(k_{ti}-q)\Theta(q-z'q_{ti}).     
\end{equation}
Another uses   $\alpha_s(q^2)$
 also for the small $z$ part (set by \verb+Ipgg=2+, \verb+ns=2+ in \ccfmupdf ): 
\begin{eqnarray}
P_{gg}(z,q,\kt ) & = & \frac{\alphasb(q^2)}{1-z} +
\frac{\alphasb(q^2)}{z} \Delta_{ns}(z,q^2,\kt)
\label{Pgg2}
\end{eqnarray}
with
\begin{equation}
\log\Delta_{ns} =  -
                  \int_0^1 \frac{dz'}{ z'}                          \int \frac{d q^2}{q^2} \alphasb(q^2)
              \Theta(k_{ti}-q)\Theta(q-z'q_{ti}).              
\end{equation}

\begin{figure}[htbp]
\centering \includegraphics[width=0.15\textwidth]{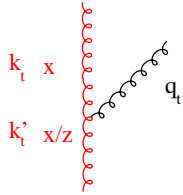}
\caption{Gluon branching}
\label{Fig:branching}
\end{figure}

In general a four-momentum  ${\bf a}$ can be written in light-cone variables as ${\bf a} = (a^+, a^-, a_T)$ with $a^+ $ and $a^-$ being the light-cone components and $a_T$ being the transverse component. The CCFM (as well as the BFKL) evolution depends only on one of the light-cone components.  Assuming that the other one can be neglected,  this leads to the condition that the virtuality of the parton propagator $a^2 = 2 a^+ a^- -a_T^2$ should be dominated by the transverse component, while the contribution from the longitudinal components is required to be small.
 The condition that $a^+a^- = 0$ leads to the so-called consistency constraint (see Fig.~\ref{Fig:branching}), which has been implemented in
different forms (set by  \verb+Ikincut=1,2,3+ in \ccfmupdf )

\begin{eqnarray}
q_t^2   & <  \frac{\kt^2}{z} & \mbox{LDC \cite{Ciafaloni:1987ur,Andersson:1995ju}}\\
q_t^2   & <   \frac{(1-z) \kt^2}{ z} &  \label{consistency-constraint}\mbox{ \cite{Kwiecinski:1996td}}\\
\kt^{'\,2} & < \frac{\kt^2}{z} &  \mbox{BFKL \cite{Kwiecinski:1996td}}
\end{eqnarray}

\subsubsection{Valence quarks}
 Using the
method of~\cite{Deak:2010gk,Deak:2011ga}
valence quarks
are included in the branching evolution    at the
 transverse-momentum dependent level  according to
\begin{eqnarray}
x{ Q_v} (x,k_t,p ) &=&  x{ Q_v}_0 (x,k_t,p ) + \int \frac{dz }{z}
\int \frac{d q^2}{ q^{2}} \Theta(p - zq)
\nonumber\\
& \times&  \Delta_s  (  p ,  zq)   
P_{qq} (z,  q , k_t) \  x{Q_v}\left(\frac{x}{z},k_t +  (1-z) q , q \right)     \;\; ,
\label{integral}
\end{eqnarray} 
where   $p$ is  the evolution  scale.  The quark splitting function  $P_{qq}$ is 
given by
\begin{eqnarray}
P_{qq}   (z,  q ,  k_t)   &=& \frac{C_F}{2\pi}  \alpha_s \left(q^2   (1-z)^2 \right)
 \frac{1+z^2}{1-z}    \;\;  .
\label{splitt}
\end{eqnarray}
In Eqs.~(\ref{integral}),(\ref{splitt})   
 the non-Sudakov form factor is not included, 
unlike the CCFM kernel  given in the appendix~B of~\cite{Catani:1989sg},
because  we only associate this   factor  with   $1/z$ terms.
  The term $x{ Q_v}_0$ in     Eq.~(\ref{integral})  is the contribution of the non-resolvable branchings
between  starting scale $q_0$ and evolution scale $p$, given by
\begin{equation}
 x{ Q_v}_0 (x,k_t,p )   =    x{ Q_v}_0 (x,k_t,q_0 )  \Delta_s  (  p ,  q_0)   \;\; ,  
  \label{Q0term}
\end{equation}
where $ \Delta_s$ is the Sudakov form factor.
\subsubsection{Sea quarks}
For a complete description of the  final states also the contribution from sea-quarks needs to be included.   We include
 splitting functions $P_{ab}$ according to
\begin{eqnarray}
P_{gg}(z)  & =  & \alphasb  \left(
 { 1 \over  {1-z_i}} - 1   + { { z_i (1 - z_i)  } \over 2 }   \right)    +
\alphasb
 \ \left(  { 1 \over z_i}   - 1   + { { z_i (1 - z_i)  } \over 2 }
\right)   \   \Delta_{ns} 
\label{splitt1} \\
P_{qg}(z)  & = &\alphasb \frac{1}{4 C_A} \left( z^2 + (1-z)^2 \right) \label{splitt2}\\
P_{gq}(z) & = & \alphasb \frac{C_F}{2C_A} \left( \frac{1 + (1-z)^2}{z}\right)\label{splitt3}\\
P_{qq}(z) & = &  \alphasb \frac{C_F}{2C_A} \left( \frac{1+z^2}{1-z}\right)\label{splitt4}
\end{eqnarray}
 with $\alphasb=C_A \alpha_s  /  \pi $, $C_A=3$ and $C_F=4/3$. 

The $g\to q\bar{q}$ splitting has been calculated in a \kt - factorized form in \cite{Catani:1994sq},
\begin{eqnarray}
P_{qg}( z, {\tilde q}, \kt)   & = &\alphasb \frac{1}{4 C_A} \left[ \frac{{\tilde q}^2}{{\tilde q}^2 + z(1-z)\ \kt^2}\right]^2 \left( z^2 + (1-z)^2 +
4 z^2 (1-z)^2 \ \frac{\kt^2}{{\tilde q }^2}
\right)
\end{eqnarray}
with ${\tilde q} = q - z\kt $,  and $q (\kt)$ being the transverse momentum of the quark (gluon).

The evolution equation for the TMD sea-quark density ${\cal S}(x,\kt,\Pmax)$, depending on
$x$, $\kt$ and the evolution variable $\Pmax$ is
(we allow a general $\kt$ dependence of the splitting functions, as proposed in appendix B of \cite{Catani:1989sg}, 
even if it is not included in eqs.(\ref{splitt1}-\ref{splitt4})),
\begin{eqnarray}
\label{usearepr1}
  {\cal S} ( x , \kt , p  ) & = &  {\cal S}_0 ( x , \kt , p  )   \nonumber \\
 & + &  \int { {dz} \over z} \int { { d q^2} \over q^2} \ \Theta   (p - z  q)  \ \Delta_s    (p , z  q) P_{qg} ( z, q, \kt)  
\   {\cal A}
 \left( { x \over z} , \kt  + (1-z) q, q \right)  \nonumber \\
 & + &  \int { {dz} \over z} \int { { d q^2} \over q^2} \
\Theta   (p - z  q)   \Delta_s    (p , z  q)  P_{qq} ( z, q, \kt)   \ {\cal S}   \left( { x \over z} , \kt  + (1-z) q, q \right) ,
\label{seaquark}
\end{eqnarray}
where $ {\cal S}_0 ( x , \kt , p  )$  is  the non-resolvable branching probability similar to Eqs.~(\ref{uglurepr2}),(\ref{Q0term}).

The evolution of the TMD gluon density including the
contribution from quarks  is given by 
\begin{eqnarray}
  {\cal A} ( x , \kt , p  ) & = &
  {\cal A}_0 ( x , \kt , p  )   \nonumber \\
  & + &
\int { {dz} \over z} \int { { d q^2} \over q^2} \
\Theta   (p - z  q)   \Delta_s    (p , z  q)  P_{gg} ( z, q, \kt)   \ {\cal A}   \left( { x \over z} , \kt  + (1-z) q, q \right) \nonumber \\
& + & \int { {dz} \over z} \int { { d q^2} \over q^2} \
\Theta   (p - z  q)   \Delta_s    (p , z  q)  P_{gq} ( z, q, \kt)   \ {\cal S}   \left( { x \over z} , \kt  + (1-z) q, q \right) .
\label{gluon}
\end{eqnarray}

\subsubsection{Monte Carlo solution of the CCFM evolution equations}

The evolution equations Eqs.(\ref{seaquark},\ref{gluon}) are  integral equations of the Fredholm type
$$
f(x) = f_0(x) + \lambda \int_a^b K(x,y) f(y)
dy
$$ and can be solved by iteration  as a Neumann series 
\begin{eqnarray}
f_1(x) & = & f_0(x) + \lambda \int_a^b
K(x,y) f_0(y) dy  \nonumber \\
f_2(x) & = & f_0(x) + \lambda \int_a^b
K(x,y_1) f_0(y_1) dy_1  +
\lambda^2 \int_a^b \int_a^b
K(x,y_1)  K(y_1,y_2)f_0(y_2) dy_2 dy_1\nonumber \\
\cdots
\end{eqnarray}
using the kernel $K(x,y)$, with the solution
\begin{equation}
f(x) =
\lim_{n\to \infty} \sum_{i=0}^n f_i(x) .
\end{equation}
\begin{figure}[htbp]
\centering \includegraphics[width=0.95\textwidth]{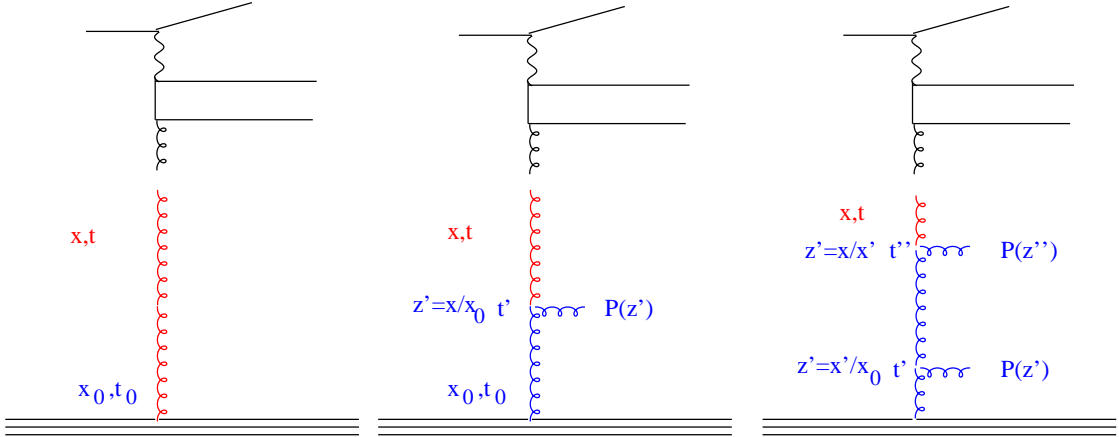}
\caption{Evolution by iteration}
  \label{Fig:evolution}
\end{figure}

Applying this to the evolution equations Eqs.(\ref{seaquark},\ref{gluon}),  we identify $f_0$ with the first term in eqs.(\ref{gluon}),
where we use for simplicity here and in the following $\Delta_s(p)=\Delta_s(p,q_0)$:

\begin{eqnarray}
{\cal A}_0(x,\kt,p) & = &   {\cal A}_0 ( x , \kt ) \Delta_s(p) .
\end{eqnarray}
The first iteration involves one branching: 
\begin{eqnarray}
{\cal A}_1(x,\kt,p) & = &   {\cal A}_0 ( x , \kt ) \Delta_s(p) \nonumber \\
&  & + \int_x^1 \frac{dz'}{z'} \int_{q_0}^p \frac{dq'^{2}}{q'^{2}}
{ \Theta   (p - z'  q') 
\frac{\Delta_s(p)}{\Delta_s(zq')}}
 \tilde{P}(z')
{ {\cal A}_0(x/z',\kt',q') }   .
\end{eqnarray}
The second iteration  involves two branchings, 
\begin{eqnarray}
{\cal A}_2(x,\kt,p) & = & {\cal A}_0 ( x , \kt ) \Delta(p) \nonumber \\
&  & + \ \int_x^1 \frac{dz'}{z'}  \int_{q_0}^p \frac{dq'^{2}}{q'^{2}}  \Theta   (p - z'  q') 
\frac{\Delta(p)}{\Delta(q')}
\tilde{P}(z') { {\cal A}_1(x/z',\kt',q') } \nonumber \\
& = &  {\cal A}_0 ( x , \kt ) \Delta(p)
+ \frac{\as}{2\pi} \int_x^1 \frac{dz'}{z'}  \int_{q_0}^p \frac{dq'^2}{q'^2}  \Theta   (p - z'  q') 
\frac{\Delta_s(p)}{\Delta_s(zq')}
\tilde{P}(z') { {\cal A}_0(x/z',\kt',q') } \nonumber \\
 &  & + \ \left(\frac{\as}{2\pi}\right)^2 \int_x^1 \frac{dz'}{z'}
 \int_{q_0}^p \frac{dq'^2}{q'^2}   \Theta   (p - z'  q') 
\frac{\Delta_s(p)}{\Delta_s(z'q')}
 \tilde{P}(z')
\nonumber \\
 &  & \times \
 \int_x^1 \frac{dz''}{z''} \int_{q_0}^p \frac{dq''}{q''}    \Theta   (p - z''  q'') 
\frac{\Delta_s(p)}{\Delta_s(z''q'')}
 \tilde{P}(z'')
{ {\cal A}_0(z''/z',\kt'',q'') } , \\
{\cal A}_3(x,\kt,p) & = & \cdots \nonumber \\
\vdots \nonumber
\label{sudakov_iteration}
\end{eqnarray}

In a Monte Carlo (MC) solution~\cite{\SMALLXC} we evolve from $q_0$ to a value $q'$ obtained from the
Sudakov factor $\Delta_s(q',q_0)$ (for a schematic visualisation of the evolution see fig.~\ref{Fig:evolution}). Note that the Sudakov factor $\Delta_s(q',q_0)$ gives the probability for evolving from
$q_0$ to $q'$ without resolvable branching.
The value $q'$ is obtained from solving for $q'$:

\begin{eqnarray}
R & = \Delta_s(q',q_0) ,
\end{eqnarray}
for a random number $R$ in $[0,1]$.

If $q' > p$ then the scale
$p$ is reached and the evolution is stopped,   and we are left with just the first
term without any resolvable branching. If $q'<p$ then we generate a branching at $q'$ according to
the splitting function $\tilde{P}(z') $, as described below, and continue the evolution using the Sudakov factor $\Delta_s(q'',q')$.
If $q'' > p$ the evolution is stopped and we are left with just one resolvable branching at $q'$. If $q''<p$
we continue the evolution as described above. This procedure is repeated until we generate
$q>p$. By this procedure we sum all kinematically allowed contributions in the series
$\sum f_i(x,p)$ and obtain an MC estimate of the parton distribution function.
 
With the Sudakov factor $\Delta_s$ and using
\begin{eqnarray*}
\frac{\partial}{\partial q'^2}\Delta_s(p,zq') = \frac{\partial}{\partial q'^2}
\frac{\Delta_s(p)}{\Delta_s(zq')} &=&
\frac{\Delta_s(p)}{\Delta_s(zq')}
\left[\frac{1}{q'^2}\right]
\int^{z_{max}} d z \tilde{P}(z)  , 
\end{eqnarray*}
we can write the first iteration of the evolution equation as  
\begin{eqnarray}
{\cal A}_1(x,\kt,p) & = &  
 {\cal A}_0 ( x , \kt, p )  \nonumber \\
 & + &  \int_x^1 \frac{dz'}{z'}  \int_{q_0}^p d\Delta_s(p,z'q')  \tilde{P}(z')  { {\cal A}_0(x/z',\kt',q') }
  \left[{\int^{z_{max}} d z \tilde{P}(z) } \right]^{-1}   .
\end{eqnarray}
The integrals can be solved by a Monte Carlo method~\cite{James:1980yn}:  $z$ is generated from

\begin{eqnarray}
\int_{z_{min}}^z  dz'\tilde{P}(z') & = & R_1 \int_{z_{min}}^{z_{max}} dz' \tilde{P}(z') ,  \label{MC2}
\end{eqnarray}
with $R_1$ being a random number in $[0,1]$,  
and  $q'$ is generated from
\begin{eqnarray}
R_2 &=&  \int_{-\infty}^x f(x')dx' = F(x) \nonumber \\
& = & \int_{zq}^p \frac{\partial}{\partial q'^2}
\left(\frac{\Delta_s(p)}{\Delta_s(zq')} \right) dq'^2 \nonumber\\
& = & \Delta_s(p,zq') \label{MC1}
\end{eqnarray}
 solving for $q'$, using $z$ from above and another random number $R_2$ in [0,1].

This completes the calculation on the first splitting. This procedure is repeated until $q'>p$ and the evolution is stopped.

With $z'$ and $q'$ selected according to the above the first iteration of the evolution equation yields
\begin{eqnarray}
x {\cal A}_1(x,\kt,p) & = &  x{\cal A}_0 ( x , \kt ) \Delta_s(p) \nonumber \\
 & + & \sum_i \tilde{P}(z'_i)  { x'_i{\cal A}_0(x'_i,k'_{t\i},q'_i) }  \left[{\int^{z_{max}} d z \tilde{P}(z) } \right]^{-1}  ,
 \end{eqnarray}
with $x'_i=x/z_i$.

\subsubsection{Normalisation of gluon and quark distributions}
 The valence quark densities are normalised so that they fulfil for every $p$ the flavor sum rule.

 The gluon and sea quark densities are normalised so that for every $p$
 \begin{eqnarray}
 \int_0^1 dx \int_0^\infty d \kt^2 x{\cal A}(x,\kt,q_0) &  =  & \int_0^1 dx \int_0^\infty d \kt^2 \left( x{\cal A}(x,\kt,p) +  x{\cal S}(x,\kt,p) \right)   .
 \end{eqnarray}

 \subsection{Computational Techniques: CCFM Grid}

When using the CCFM evolution in a fit program to determine the starting distribution ${\cal A}_0 (x)$, a full MC solution~\cite{\SMALLXC} is no longer suitable, since it is time consuming and suffers from numerical fluctuations. Instead a convolution method introduced   in~\cite{Jung:2012hy,Hautmann:2013tba} is used. The
kernel $ \tilde {\cal A}\left(x'',\kt,\Pmax\right) $ is determined once from the Monte Carlo  solution of the CCFM evolution equation, and then folded with the non-perturbative starting distribution ${\cal A}_0 (x)$,
\begin{eqnarray}
x {\cal A}(x,\kt,\Pmax) &= &x\int dx' \int dx'' {\cal A}_0 (x') \tilde{\cal A}\left(x'',\kt,\Pmax\right)
 \delta(x'
x'' - x)
\nonumber 
\\
& = & \int dx' {{\cal A}_0 (x') } 
\cdot \frac{x}{x'} \ { \tilde{\cal A}\left(\frac{x}{x'},\kt,\Pmax\right) } .
\end{eqnarray}
The kernel  $\tilde{\cal A}$ incorporates all of  the dynamics of the evolution, including   Sudakov form factors and splitting functions.  It  is determined on a grid of $50\otimes50\otimes50$ bins  in $ x,  \kt,  \Pmax$.   The binning in the  grid is  logarithmic,  except for  the longitudinal variable   $x$ where  we use 40 bins in logarithmic spacing below 0.1, and 10 bins in linear spacing above 0.1.

Using this method, the complete coupled evolution of gluon and sea quarks is more complicated, since it is no
longer a simple convolution of the kernel with the starting distribution.
To simplify the approach, here we allow only for one  species of partons at the starting scale,  either gluons or sea-quarks. During evolution the other species will be generated. This approach, while convenient  for QCD fits, has the feature that sea-quarks,  in the case of gluons only at $q_0$,  are generated with perturbative transverse momenta ($\kt > k_{t\ cut}$), without contribution from the soft (non-perturbative) region.

\subsection{Functional Forms for starting distribution}
\subsubsection{Standard parametrisation}
For the starting distribution  ${\cal A}_0$,  at the starting scale $q_0$,  the following form is used:
\begin{eqnarray}
x{\cal A}_0(x,\kt,q_{0}) = A_1 x^{-A_2} \cdot (1 -x)^{A_3}\left( 1 -A_4 x
+A_5 \sqrt{x}  +A_6 x^2 \right)
   \exp[ - k_t^2 / \sigma^2 ]  \;\; ,
\label{a0-5par}
\end{eqnarray}
with $ \sigma^2  =  q_{0}^2 / 2 $ and free parameters $A_1, \dots, A_6 $.

Valence quarks are treated  using the method
of~\cite{Deak:2010gk,Deak:2011ga,Hautmann:2013tba} with starting distributions  at scale $q_{0}$
parameterized using standard collinear pdfs (set by \verb+Ipdf+ in \ccfmupdf )   as
\begin{equation}
    x{ Q_v}_0 (x,k_t,q_{0} ) =    x{ Q_v}_{\rm{coll. pdf}} (x,q_{0} ) \
    \exp[ - k_t^2 / \sigma^2 ]   \;\;   .   
\label{gauss}
\end{equation}
with  $ \sigma^2  =  q_{0}^2 / 2 $.
At every scale $p$  the flavor sum rule is fulfilled for valence quarks.

\subsubsection{Saturation ansatz}
A saturation ansatz for the starting distribution  ${\cal A}_0$  at  scale $q_0$  is available,
following the parameterisation of the
saturation model by Eq.(18) of ~\cite{wuesthoff_golec-biernat},
\begin{eqnarray}
x {\cal A}_{sat} = \frac{1}{\alpha_s} \frac{3 \sigma_0}{4 \pi^2} R_0^2(x)\kt^2 \exp{\left( -R_0^2(x) \kt^2 \right)} ,
\end{eqnarray}
with $R_0^2(x) = (x/x_0)^\lambda$. The free parameters are $\sigma_0 = A_2$, $\lambda = A_3$, $x_0 = A_4$ and
$\alpha_s=A_5$.
In order to be able to use this type of parameterisation over the full $x$ range,  an additional factor of $(1-x)^{A_6}$ (see \cite{Grinyuk:2013tt}) is applied.
\subsection{Plotting TMDs}
A simple plot program is included in the package. For a graphical web interface use \TMDplotter ~\cite{tmdplotter}.
\subsection{Application}

The evolution of the TMD gluon density has been used to perform fits
to the DIS precision data~\cite{Aaron:2009aa,Abramowicz:1900rp}, as
described in detail in \cite{Hautmann:2013tba}.

\section{Description of the program components}

\subsection{Program history}
\begin{tiny}

\begin{verbatim}
*________________________________________________________________________
uPDFevolv 
*            Version  10000
*            first public release
*________________________________________________________________________
\end{verbatim}
\end{tiny}
\subsection{Subroutines and functions}

The source code of \ccfmupdf\ and this manual can be found under:\\
\verb+https://updfevolv.hepforge.org/+

\begin{defl}{123456789012345678}
\item[{\tt      sminit}] to initialise
\item[{\tt      sminfn}] to generate starting distributions in $x$ and $k_t$
\item[{\tt      smbran}] to simulate perturbative branchings
\item[{\tt      splittgg}] to generate $g\to g g$  splitting via $P_{gg}$
\item[{\tt      splittgq}] to generate $q \to g q$ splitting via $P_{gq}$
\item[{\tt      splittqg}] to generate $g \to q\bar{q} $ splitting via $P_{qg}$
\item[{\tt      splittqq}] to generate $q \to q g $ splitting via $P_{qq}$
\item[{\tt      szvalnew}] to calculate $z$ values for $g\to gg$ splitting
\item[{\tt      smqtem}] to generate $t$ from the corresponding Sudakov factor
\item[{\tt      updfgrid}] to build, fill and normalise the updf grid.
\item[{\tt      asbmy(kt)}]    to calculate $\frac{ C_A } { \pi}\alpha_s(k_t)$
\\

\item[{Utility routines:}]
\item[{\tt evolve tmd}] Main routine to perform CCFM evolution
\item[{\tt updfread}] example program to read and plot the results
\item[{\tt  gadap}] 1-dimensional Gauss integration routine	
\item[{\tt  gadap2}] 2-dimensional Gauss integration routine
\item[{\tt  divdif}]      linear interpolation routine  (CERNLIB)
\item[{\tt      ranlux}]  Random number generator \verb+RANLUX+ (CERNLIB)
\end{defl}

\subsection{Parameter in steering files}
\begin{defl}{123456789012345678}
\item[]
\item[{\tt  'updf-grid.dat'}] name of the grid file
\item[{\tt  oneLoop = 0}] to select {\it all loop} CCFM or {\it one loop} DGLAP type evolution
\item[{\tt  saturation = 0}] to select standard or saturated initial condition
\item[{\tt  Ipdf = 60500}] LHApdf set name for collinear valence quark starting distribution
\item[{\tt  Itarget = 2212}] hadron target ID (2212=proton)
\item[{\tt  Iglu = 1}]  for gluon only evolution
\item[{\tt  Ipgg = 1}] parameter for $P_{gg}$ splitting function
\item[{\tt  ns =  1}] parameter for treatment of non-sudakov form factor
\item[{\tt  ikincut = 2}] flag for consistency constraint
\item[{\tt  Qg = 2.2}] starting value $q_0$ for perturbative evolution
\item[{\tt  QCDlam = 0.20}] value for $\Lambda_{qcd}$
\item[{\tt  A1,..., A6}]  values for starting distribution;  meaning depends on  whether standard or saturation ansatz is used.
  
\end{defl}

\section{Example Program}
\begin{tiny}
\begin{verbatim}
      Program ccfm_uPDF

      Include 'SMallx.inc'
      Integer Iev
C--- event common block
      Integer NMXHEP,NEVHEP,NHEP,ISTHEP,IDHEP,JMOHEP,JDAHEP
      Double Precision PHEP,VHEP,EVWGT
C---Event weight
      COMMON/HEPWGT/EVWGT
      Integer nobran,ikincut
      Common/myvar/nobran,ikincut
      Double Precision Qbarmy,Qbar_min,Qbar_max
      Common/mglubran/Qbarmy,Qbar_min,Qbar_max
      Integer neve
      Common/myevt/neve
      Integer nloop
      Common/myloop/nloop
      Double Precision x3lmin,x3lmax,x3ldif
      Integer Nbp
      Parameter (Nbp=50)
      Double Precision X3(0:Nbp+1)
      Double Precision X3M(0:Nbp)
      Double Precision x3b(0:Nbp)
      Common/gridtt/x3m

      Integer ng_max
      Integer nrglu
      Common/mynrglu/nrglu
      Integer nmax,i,nx3,kev,ic
      Integer Ipgg,ns_sel
      Double Precision scal
      Common/Pggsel/Ipgg,ns_sel,scal
      Double Precision Qgmin
      Character *72 TXT
      CHARACTER   FILNAME*132,testNAME*132
      Common/gludatf/filname
      Double Precision Xnorm
      Common/smnorm/ Xnorm
      Logical pdflib,quark,gluon,photon,saturation
      Common /SMbran2/pdflib,quark,gluon,photon,saturation
      Integer Ioneloop,Itarget,Iglu,Isaturation
      Integer Ipdf
      Common/pdf/Ipdf
      Integer iparton
      Common /SMquark/iparton
      Character *15 char
      Double Precision BB
      Common /splitting/ BB
      Double precision ininorm(-6:6)
      Common/smininorm/ininorm
    
      Integer IRR
      Couble precision au
      Logical first
      Common/f2fit/au(50),first      
*
      Read(5,*) filname
      Write(6,*) ' output file ',filname
      xnorm = 1.
      Read(5,101) TXT
      Read(txt,1005) char,Ioneloop
      Write(6,*) txt,char,Ioneloop
1005  format(a10,I8)
      Read(5,101) TXT
      Read(txt,1010) char,Isaturation
1010  format(a14,I8)
      Write(6,*) txt,char,Isaturation
      Read(5,101) TXT
      Read(txt,1006) char,Ipdf
      Write(6,*) txt,char,Ipdf
1006  format(a7,I8)
      Read(5,101) TXT
      Read(txt,1007) char,Itarget
      Write(6,*) txt,char,Itarget
1007  format(a10,I8)
      Read(5,101) TXT
      Read(txt,1008) char,Iglu
      Write(6,*) txt,char,Iglu
1008  format(a7,I8)
      Read(5,101) TXT
  101 Format(A72)
      Read(txt,1000) char,Ipgg
      Write(6,*) txt,char,Ipgg
1000  format(a7,I8)
      Read(5,101) TXT
      Read(txt,1001) char,ns_sel
      Write(6,*) txt,char,ns_sel
1001  format(a5,I8)
      Read(5,101) TXT
      Read(txt,1011) char,ikincut
      Write(6,*) txt,char,ikincut
1011  format(a10,I8)
      Read(5,101) TXT
      Read(txt,1002) char,Qg
      Write(6,*) txt,char,Qg
1002  format(a5,F16.8)
      Read(5,101) TXT
      Read(txt,1003) char,Qs
      Write(6,*) txt,char,Qs
1003  format(a5,F16.8)
      Read(5,101) TXT
      Read(txt,1018) char,QCDlam
      Write(6,*) txt,char,QCDlam
1018  format(a9,F16.8)
      Read(5,101) TXT
      Read(txt,1003) char,AU(1)
      Read(5,101) TXT
      Read(txt,1003) char,AU(2)
      Read(5,101) TXT
      Read(txt,1003) char,AU(3)
      Read(5,101) TXT
      Read(txt,1003) char,AU(4)
      Read(5,101) TXT
      Read(txt,1003) char,AU(5)
      Read(5,101) TXT
      Read(txt,1003) char,AU(6)
     
      If(iglu.ne.0) then
          gluon = .true.
          else
          gluon = .false.
        Endif
      If(Ioneloop.eq.1) then
          onel = .true.
          else
          onel=.false.
      Endif
      If(Isaturation.eq.1) then
          saturation = .true.
          else
          saturation=.false.
      Endif
      If(iglu.eq.0) then
          Read(50,101) TXT 
          Read(txt,1009) char,Iparton
          Write(6,*) txt,char,Iparton
          Write(6,*) txt
1009  Format(a9,I8)
      Endif
      Close(50)

C---Initialize run
      Call SMinit
      neve = 0
      nmax =nev
      Xini = 0.
      Write(6,*) ' output file ',filname
      Write(6,*) ' selection Ipgg = ',Ipgg,' ns_sel = ',ns_sel
      Write(6,*) ' Qg = ',Qg,' Qs = ',Qs,' Xnorm = ',Xnorm
      Write(6,*) ' LHAPDFLIB for val quark Ipdf = ',Ipdf
      Write(6,*) ' Itarget = ',Itarget
      Write(6,*) ' BB = ',BB

      Qgmin = max(Qg-0.5d0,QCDlam)
      Qgmin = max(Qg,QCDlam)
      x3lmin = log(Qgmin) 
      x3lmax = log(qmax)
      x3ldif = (x3lmax-x3lmin)/Real(Nbp)
      Do I=0,Nbp+1
          x3(I) = exp(x3lmin + x3ldif*Real(I))
      Enddo
      Do I=0,Nbp
          x3m(i) = (x3(i) + x3(i+1))/2.
          x3b(i) = x3(i+1) - x3(i)
      Enddo
      Nx3 = -1
C---Initialize analysis
      Xini=0.
      Xfin=0.
      Call updfgrid(1)
      Nx3 = 49
600  Nx3 = Nx3 + 1
      write(6,*) ' ng_max = ',ng_max,' at nx3-1 ',nx3-1
      IF(Nx3.gt.Nbp) Then
          write(6,*) ' evolve_tmd: Nx3 gt Npb -> Program stopped'
          stop
      Endif
      Qbarmy = x3m(Nx3)
      nloop = Nx3
      ng_max = 0

      Write(6,*) ' Qbar_my  ',Qbarmy
     
      Do Iparton=0,2
           Write(6,*) ' evolving parton ID: ',Iparton
           If(iparton.eq.0) then
                gluon=.true.
                else
               gluon=.false.
           Endif
          Xini = 0
          Do I=1,nev
C---Initialize event
              Call SMinfn

C---gluon branching process
              Call smbran
              Xini = Xini + x0
              If(ng_max.le.nrglu) ng_max=nrglu
              If(wt.NE.0.0) then
                  kev=i
                  if (kev.gt.0) ic=100000
                  if (mod(kev,ic).eq.0) write(6,*) ' event ',kev,nev,' loop P_max ',nloop,Qbarmy
              Endif
          Enddo
      Enddo
C---Terminate analysis
      call updfgrid(3)
      Stop
80  Write(6,*) ' steering file ccfm_updf not found '
      stop   
      End

            
\end{verbatim}
\end{tiny}
\section{Program Installation}
 \ccfmupdf\ follows the standard AUTOMAKE
convention. To install the program, do the following
\begin{tiny}
\begin{verbatim}
1) Get the source

tar xvfz uPDFevolv-XXXX.tar.gz
cd uPDFevolv-XXXX

2) Generate the Makefiles (do not use shared libraries)
./configure

3) Compile the binary
make

4) Install the executable
make install

4) The executable is in bin

run it with:
bin/updf_evolve < steer_gluon-JH-2013-set2

plot the result with:

bin/updfread

\end{verbatim}
\end{tiny}

 \section{Acknowledgments}
We are very grateful to Bryan Webber for careful reading of the manuscript and clarifying comments.

\bibliographystyle{heralhc}
\raggedright
\bibliography{ref}

\end{document}